\documentclass[a4paper]{report}
\usepackage[utf8]{inputenc}
\usepackage[T1]{fontenc}
\usepackage{abc}
\usepackage{amsmath,amssymb,array}
\usepackage{booktabs}

\begin{document}

\sectionhead{}

\begin{article}
  
\title{The Causal Effects for a Causal Loglinear Model}
\author{by Gloria Gheno }
\address{\\
  European Centre for Living Technology \\
  Ca' Minich, San Marco 2940, 30124 Venice, ITALY \\
  and Department of Environmental Sciences, Informatics and Statistics\\
  Ca’ Foscari University of Venice \\
  Dorsoduro 2137, 30123 Venice, ITALY\\}
\email{gloria.gheno@unive.it}
\maketitle

\abstract{The analysis of the causality is important in many fields of research. I propose a causal theory  to obtain the causal effects in a causal loglinear model. It calculates them using the odds ratio and Pearl's causal theory. The effects are calculated distinguishing between a simple mediation model (model without the multiplicative interaction effect) and a mediation model  with the multiplicative interaction effect. In both models it is possible also to analyze the cell effect, which is a new interaction effect. Then in a causal loglinear model there are three interaction effects: multiplicative interaction effect, additive interaction effect and cell effect}

\section{Introduction}

The analysis of the causality is important in many fields of research, for example in economics and in social sciences, because the analyst seeks to understand the mechanisms of the analyzed phenomena  using the relations among the variables (i.e. the relations cause-effect, where some variables are the causes, other variables the effects). These variables can influence directly,  indirectly or in both ways other variables. The set of all effects which influence a variable is called total effect. The direct effect is the effect of a variable on another variable without any intervening variables, while the indirect effect  is the effect of a variable on another variable considering only the effect through the intervention of  other variables, called mediators.  \cite{Wright21} defines a diagram for the causal relations, which he calls "path diagram". In the path diagram, the direct causal relation between 2 variables is represented by an arrow which goes from the influencing variable  to the influenced variable . If two variables are not connected, then there is not direct causal relation between them. The correlation between two variables is represented by a double arrow. To explain better the direct, indirect and total effects then I use the path diagram represented in Figure 1: the arrow which goes from X to Y represents the direct effect of X on Y,  the two arrows which go from X and Z to Z and Y represent the  indirect effect of X on Y through Z and the arrow which goes from Z to Y represents the direct effect of Z on Y. Then the indirect effect is the effect of X on Y mediated by Z.  An analyst, then, who is interested in the variable Y, will be interested to understand what affects Y and then he will study the direct, indirect and total effects. 

It is possible to complicate these effects by introducing the concept of interaction. The interaction occurs when the effect of one cause-variable may depend in some way on the presence or absence of another cause-variable. In literature the interaction effect can be measured on the additive or multiplicative scale and in many case induces that the effect of one variable on another varies by levels of a third and vice versa. Figure 2 shows the path diagram of the interaction, where X and Z influence directly Y but also their joint effect XZ influences Y.   Both interaction effects can be present in a model. A problem of using the loglinear models is the inability to calculate all these effects and this can be considered  its limitation. In this paper I propose a causal theory which provides a method for calculating such effects in a loglinear model. 

\begin{figure}[t]
\centering
\begin{minipage}[c]{.57\textwidth}
\centering
\includegraphics[width=.57\textwidth]{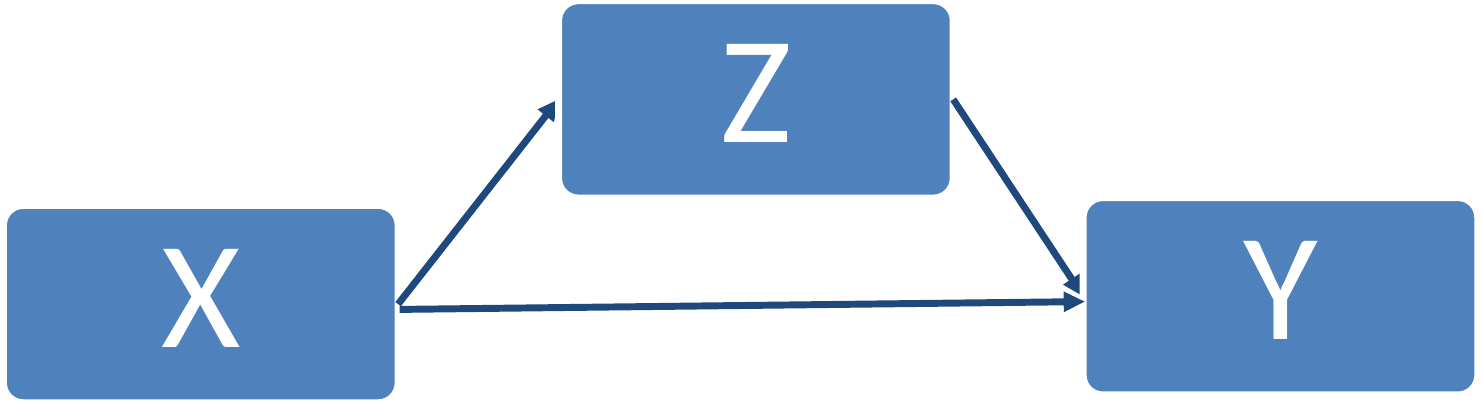}
\caption{Simple mediation model}
\end{minipage}
\hspace{0.1 mm}
\begin{minipage}[c]{.38\textwidth}
\centering
\includegraphics[width=.38\textwidth]{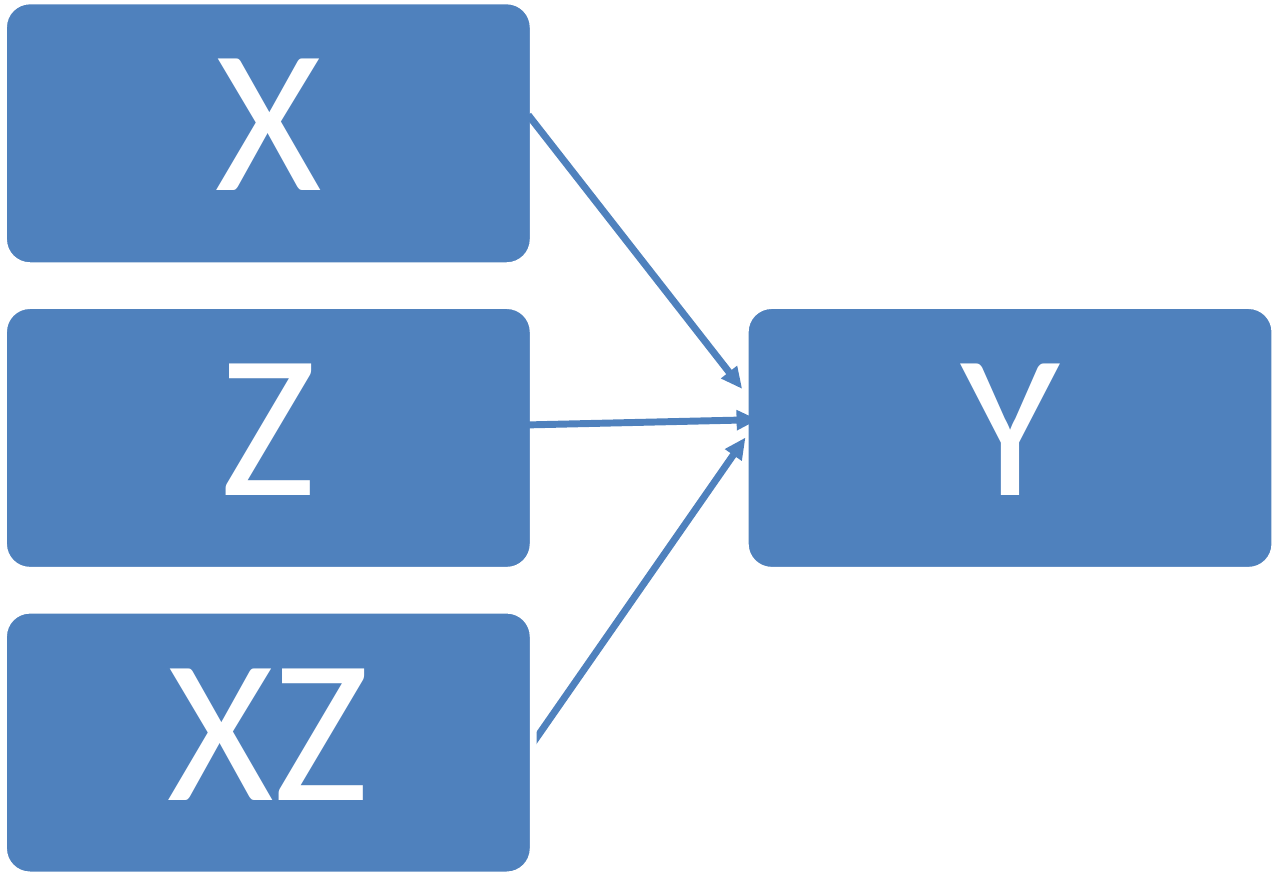}
\caption{Simple interaction model}
\end{minipage}

\end{figure}

\section{Causal loglinear model with or without multiplicative interaction}

Before introducing the method to calculate the effects, I explain the transition from a loglinear model to a causal loglinear model which represents a loglinear model where the variables have a causal role, i.e. for example X becomes the cause and Y the effect.  \cite{Ver96}, indeed, distinguishes the loglinear models in these 2 models, which he calls respectively loglinear models and causal loglinear models. The loglinear model describes the observed frequencies, it doesn't distinguish between dependent and independent variables and it measures the strength of the association among  variables. The causal loglinear model, introduced by  \cite{Goodman73} and also called "modified path analysis approach",  is  a loglinear model which considers a causal order of the variables a priori. This model, as written by \cite{Ver05},  consists of specifying a "recursive" system of logit models. In this system the variable, which appears as dependent  in a particular logit equation, may appear as one of the independent variables in one of the next equations.  For simplicity, I consider a model with 3  categorical variables, X, Z and Y. The joint probability in multiplicative form is 

\begin{equation*}
\begin{array}{rl}
P(X=x, Z=z, Y=y)  =& \pi^{X=x,Z=z,Y=y}\\
= &\eta \mu^{X=x} \mu^{Y=y} \mu^{Z=z} \mu^{X=x,Y=y} \mu^{X=x,Z=z} \mu^{Z=z,Y=y} \\
& \mu^{X=x, Z=z, Y=y}\\
\end{array}
\end{equation*}
\\
which can be written also in additive form
\begin{equation*}
\begin{array}{rl}
\log{P(X=x, Z=z, Y=y)}  = &\log{\eta}+ \log{ \mu^{X=x}}+ \log{ \mu^{Y=y}}\\
 & + \log{ \mu^{Z=z}}+ \log{ \mu^{X=x,Y=y}} +\log{ \mu^{X=x,Z=z}} +\log{ \mu^{Z=z,Y=y}} \\
& + \log{\mu^{X=x, Z=z, Y=y}}\\
\end{array}
\end{equation*}
\\
Now I suppose that X, Z and Y are binary ( 0 or 1), and I consider the dummy code, that is:
\begin{equation*}
\mu^{Y=0}=\mu^{X=0}=\mu^{Z=0}=1
\end{equation*}
\begin{equation*}
\mu^{Z=0,Y=0}=\mu^{X=0, Y=0}=\mu^{Z=0,Y=1}=\mu^{X=0, Y=1}=\mu^{Z=1, Y=0}=\mu^{X=1, Y=0}=1
\end{equation*}
\begin{equation*}
\mu^{X=0, Z=0}=\mu^{ X=0,Z=1}=\mu^{ X=1,Z=0}=1
\end{equation*}
\begin{equation*}
\mu^{X=0,Z=0,Y=i}=\mu^{X=0,Z=1,Y=i}=\mu^{X=1,Z=0,Y=i}=\mu^{X=1,Z=1,Y=0}=1 \quad  \mbox{with} \: i=0,1
\end{equation*}
\\
The joint probability is shown in table 1. 
\begin{table}[b]
\begin{minipage}[b]{0.9\textwidth}
\centering
\begin{tabular}{||l|l|l|l||}
\hline
x & z & y & $\pi^{X=x,Z=z,Y=y}$ \\
\hline
0 & 0 & 0 & $\eta$  \\
0 & 0 & 1 & $\eta \mu^{Y=1}$  \\
0 & 1 & 0 & $\eta \mu^{Z=1}$  \\
0 & 1 & 1 & $\eta \mu^{Y=1} \mu^{Z=1} \mu^{Z=1,Y=1}$  \\
1 & 0 & 0 & $\eta \mu^{X=1}$  \\
1 & 0 & 1 & $\eta \mu^{X=1} \mu^{Y=1} \mu^{X=1,Y=1}$  \\
1 & 1 & 0 & $\eta \mu^{X=1} \mu^{Z=1} \mu^{X=1,Z=1}$  \\
1 & 1 & 1 &  $ \eta \mu^{X=1} \mu^{Y=1} \mu^{Z=1} \mu^{X=1,Y=1} $ \\
& & & $\mu^{X=1,Z=1} \mu^{ Z=1,Y=1} \mu^{X=1, Z=1,Y=1}$\\
\hline
\end{tabular}
\caption{The joint probability}
\end{minipage}
\end{table}
\begin{table}[t]

\begin{minipage}[b]{0.9\textwidth}
\centering
\begin{tabular}{||l|l|l||}
\hline
& \footnotesize{$Y=0$} & \footnotesize{$Y=1$}  \\
\hline
\footnotesize{$X=0 $} & \footnotesize{$\eta \mu^{Z=1}$} & \footnotesize{$\eta \mu^{Y=1} \mu^{Z=1}$}  \\
\hline
\footnotesize{$X=1$} & \footnotesize{$\eta \mu^{X=1} \mu^{Z=1}$} & \footnotesize{$ \eta \mu^{X=1} $}\\
&\footnotesize{$ \mu^{Z=1,Y=1}$}& \footnotesize{$\mu^{Y=1}$} \footnotesize{$ \mu^{Z=1} \mu^{X=1,Y=1}$ }\\
& & \footnotesize{$\mu^{X=1,Z=1}$} \footnotesize{$ \mu^{ Z=1,Y=1}$} \footnotesize{$\mu^{X=1, Z=1, Y=1}$}\\
\hline
\end{tabular}
\caption{Marginal table $XY$ given $Z=1$}
\end{minipage}
\end{table}

Now I consider the model of Figure 1, which gives a priori informations on the causal order. To consider the model of Figure 1 in loglinear terms, however, I must suppose that the three-interaction term is equal to 1 because, if  it is present, it introduces the causal multiplicative  interaction term of X and Z on Y (Figure 2). The presence or absence of this parameter, indeed, brings about the presence or absence of the multiplicative interaction. The multiplicative interaction is measured  calculating the  odds ratios:

\begin{equation*}
\frac{\frac{Pr(Y=1|X=1,Z=1)}{1-Pr(Y=1|X=1,Z=1)}\frac{1-Pr(Y=1|X=0,Z=0)}{Pr(Y=1|X=0,Z=0)}}
{\frac{Pr(Y=1|X=0,Z=1)}{1-Pr(Y=1|X=0,Z=1)}\frac{Pr(Y=1|X=1,Z=0)}{1-Pr(Y=1|X=1,Z=0)}}
\end{equation*}
If this ratio is equal to 1, then there is not multiplicative effect, and this occurs only if $\mu^{X=1,Z=1,Y=1}$ 
is equal to 1 or $\log{\mu^{X=1,Z=1,Y=1}}$ is equal to 0. This interaction effect can be interpreted as the interaction effect of the traditional linear model.  Following the probability structure proposed by \cite{Goodman73}, the causal model of Figure 1 can be written $P(X,Z,Y)=P(Y|Z,X) P(Z|X)P(X)$: the causal model is, then, a decomposition of the joint probability into conditional probabilities.

Now I consider the relation between the loglinear model and the causal loglinear model.  I calculate the conditional probabilities using the joint probability and the marginal probabilities. For example the conditional probability of Y=1 given X=1, Z=1, i.e. $\pi^{Y=1|X=1, Z=1}$, is calculated using  table 2 and constraining the three-interaction term equal to 1: 

\begin{equation*}
\pi^{Y=1|X=1, Z=1}=\frac{\pi^{Y=1,X=1|Z=1}}{\pi^{Y=1,X=1|Z=1}+ \pi^{Y=0,X=1|Z=1}} = \frac{\mu^{Y=1} \mu^{X=1,Y=1} \mu^{ Z=1,Y=1}} {1+ \mu^{Y=1} \mu^{X=1,Y=1} \mu^{ Z=1,Y=1}}
\end{equation*}
\\
 For simplicity, I write this conditional probability as 

\begin{equation*}
\eta^{Y| X=1, Z=1}\mu^{Y=1} \mu^{ X=1,Y=1} \mu^{ Z=1,Y=1}
\end{equation*}
\\
which I call causal form, where

\begin{equation*}
\eta^{Y|X=1,Z=1}= \frac {1}{ 1+ \mu^{Y=1} \mu^{X=1,Y=1} \mu^{ Z=1, Y=1}}
\end{equation*}
\\
can be seen as a normalization factor. This can be proved recalling that the sum of conditional probabilities $P(Y=1| X=1,Z=1)$ and $P(Y=0| X=1,Z=1)$ is equal to 1.  If I write the probabilities in causal form I have  

\begin{equation*}
\left \{ \begin{array}{l}
P(Y=1| X=1,Z=1) = \eta^{Y| X=1, Z=1}\mu^{Y=1}$ $ \mu^{ X=1,Y=1} \mu^{ Z=1,Y=1}\\
P(Y=0|X=1,Z=1)=\eta^{Y| X=1, Z=1}\mu^{Y=0} \mu^{ X=1,Y=0} \mu^{ Z=1,Y=0}=\eta^{Y| X=1, Z=1}\\
\end{array} \right.
\end{equation*}
\\
 where, in this case, I do not assume particular values for $\eta^{Y|X=x,Z=z}$. The sum of conditional probabilities is equal to  $\eta^{Y| X=1, Z=1} (1+ \mu^{Y=1} \mu^{X=1,Y=1} \mu^{Y=1, Z=1})$. Recalling that this sum must be equal to 1, I obtain that $\eta^{Y|X=1,Z=1}$ is equal to $(1+ \mu^{Y=1} \mu^{X=1,Y=1} \mu^{ Z=1,Y=1})^{-1}$  which is exactly the value which I obtain rewriting the conditional probability in causal form. For this reason, $\eta^{Y|X=x,Z=z}$ can be seen as a normalization factor. The conditional probability  $P(Z=z|X=x)$ is calculated using the table $XZ$. Then I write the marginal  probability of X and the conditional probabilities of Z and Y  in causal form:

\begin{equation}
\pi^{X=x}=\eta_c^{X} \mu_c^{X=x}
\end{equation}
\begin{equation}
\pi^{Z=z| X=x}= \eta_c^{Z|X=x} \mu_c^{Z=z} \mu_c^{ X=x,Z=z}
\end{equation}
\begin{equation}
\pi^{Y=y|X=x, Z=z}=\eta^{Y| X=x, Z=z} \mu^{Y=y} \mu^{ X=x,Y=y} \mu^{ Z=z, Y=y}
\end{equation}
\\
where for example the ratio between the causal one-effect parameter $\mu_c^{Z=1}$ and the no causal one-effect parameter $\mu^{Z=1}$ is $\eta^{Y|X=0,Z=0}/ \eta^{Y|X=0,Z=1}$ and the ratio between the causal two-effects parameter $\mu_c^{Z=1,X=1}$ and the no causal two-effects parameter $\mu^{X=1, Z=1}$ is 

\begin{equation*}
\frac{\eta^{Y|X=1,Z=0} \eta^{Y|X=0, Z=1}} {\eta^{Y|X=0,Z=0}\eta^{Y|X=1,Z=1}}
\end{equation*}
\\
The  causal normalization factors $\eta^{Y|X=x,Z=z}$ and $\eta_c^{Z|X=x}$ are calculated so: 

\begin{equation*}
\left \{ \begin{array}{l}
\eta^{Y|X=0,Z=0}= \frac{1}{1+\mu^{Y=1}} \\
\eta^{Y|X=1,Z=0}=\frac{1} {1+\mu^{Y=1} \mu^{X=1,Y=1}} \\
\eta^{Y|X=0,Z=1}= \frac{1} {1+\mu^{Y=1} \mu^{Z=1,Y=1}} \\
\eta^{Y|X=1,Z=1}= \frac{1} {1+\mu^{Y=1} \mu^{Z=1,Y=1} \mu^{X=1,Y=1}}\\
\eta_c^{Z| X=0}= \frac{1} {1+\mu_c^{Z=1}} \\
 \eta_c^{Z| X=1}= \frac{1}{1+\mu_c^{Z=1} \mu_c^{X=1,Z=1}}
 \end{array} \right.
\end{equation*}

\section{Odds ratio and causal loglinear model}

\begin{table}[t]\begin{minipage}[b]{0.4\textwidth}
\centering
\begin{tabular}{||l|l|l||}
\hline
 & \footnotesize{$Y=0$} & \footnotesize{$Y=1$}  \\
\hline
\footnotesize{$X=0$}  & \footnotesize{$\eta$} & \footnotesize{$\eta \mu^{Y=1}$} \\
\hline
\footnotesize{$X=1$} & \footnotesize{$\eta \mu^{X=1}$}  &  \footnotesize{$\eta \mu^{X=1} \mu^{Y=1} $ }\\ 
& & \footnotesize{$\mu^{X=1,Y=1}$}  \\ 
\hline
\end{tabular}
\caption{Marginal table $XY$ given $Z=0$}
\end{minipage}
\begin{minipage}[b]{0.5\textwidth}
\centering
\begin{tabular}{||l|l|l||}
\hline
 & \footnotesize{ $Y=0$} & \footnotesize{$Y=1$}  \\
\hline
\footnotesize{$X=0$} & \footnotesize{$\eta ( 1 +\mu^{Z=1})$} & \footnotesize{$\eta \mu^{Y=1} $}\\
& & \footnotesize{ $(1+ \mu^{Z=1} \mu^{ Z=1, Y=1})$ } \\
\hline
\footnotesize{$X=1$} & \footnotesize{$\eta \mu^{X=1} $}  & \footnotesize{$\eta \mu^{X=1} $} \footnotesize{$\mu^{Y=1} \mu^{ X=1,Y=1} $}\\
& \footnotesize{$(1+ \mu^{Z=1} $} & \footnotesize{$(1+ \mu^{Z=1} \mu^{X=1,Z=1}$}\\
& \footnotesize{$\mu^{Z=1, X=1})$}& \footnotesize{$ \mu^{ Z=1, Y=1}) $}\\
\hline
\end{tabular}
\caption{Marginal table $XY$ }
\end{minipage}
\end{table}

In the loglinear model, the causal effects are considered in partial way and for this reason, a true causal analysis is not made.  If I consider the causal model of Figure 1, \cite{Ber09} calculate the total effect  by the  marginal table $XY$ (table 4) and the direct effect by the 2 marginal tables $XY$ given $Z=z$ (tables 2 and 3) using the odds ratio.  The odds ratio  describes the relationship among binary variables; if the variables are categorical, it is necessary a transformation in binary variables to use them. For example if I want analyze the relation between X and Y, which are categorical variables with 5 categories, I transform them in binary variables: the transformed X and Y are equal to 1 if their original value is 5, 0 otherwise. The relationships considered  by the odds ratio can be associative or causal (\cite{Zhang}) in the first type the relation is measured using the actual response variable, while in the second using the potential response. If the two types of odds ratio are different, this is due to the influence of  a third variable called confounding variable (\cite{Zhang}; \cite{Szumilas}). This confounding variable is causally linked to  the response variable but it is not related causally to other cause or it is linked causally but it is not a mediator variable (\cite{Szumilas}) for example if X and Z influence Y, and X and Z are correlated (link which is not of causal type), Z is a confoundering variable of the relation between X and Y. Then in a simple mediation model without confounders,  the total effect (TE) and the direct effect used in the loglinear literature (LDE)  are given by the following formulas:

\begin{equation}
OR_{x,x'}^{TE} =\frac{P(Y|X=x')}{1- P(Y|X=x')} \frac{1- P(Y|X=x)}{P(Y|X=x)} 
\end{equation}

\begin{equation}
OR_{x,x'}^{LDE}(Z)=\frac{P(Y|X=x',Z=z)}{1-P(Y|X=x',Z=z)} \frac{1-P(Y|X=x,Z=z)}{ P(Y|X=x,Z=z)}
\end{equation}
\\
where the subscript $x,x'$ indicates that the odds ratio measures the effect of the variation of X from x to $x '$. I note that they coincide with the definitions of total effect and controlled direct effect proposed by \cite{Pearl01, Pearl09, Pearl12}. I remember however that Pearl never uses the odds ratio to calculate the effects,  but prefers to calculate them using the conditional moments.  For this reason, I propose a causal analysis for the loglinear models,  applying Pearl's theory to the odds ratio. Using the dummy code, the total effect is equal to

\begin{equation*}
\begin{split}
\underbrace{\mu^{Y=1,X=1}}_{\mbox{direct effect}} \left \{ \left[ \frac{ \eta^{Y|X=0,Z=0} + \mu_c^{Z=1} \eta^{Y|X=0,Z=1}}{\eta^{Y|X=1,Z=0} +\mu_c^{Z=1} \mu_c^{X=1,Z=1} \eta^{Y|X=1,Z=1}} \right]\right. \\
\left. \left [ \frac{ \eta^{Y|X=0, Z=0} + \mu_c^{Z=1} \mu^{Z=1,Y=1} \eta^{Y|X=0,Z=1}}{\eta^{Y|X=1,Z=0} +\mu_c^{Z=1} \mu_c^{X=1,Z=1}  \mu^{Z=1,Y=1} \eta^{Y|X=1,Z=1}} \right]^{-1}\right \}\\
\end{split}
\end{equation*}
\\
and the direct effect used in the loglinear literature is equal  always to the causal two-effects parameter $\mu^{X=1,Y=1}$, i.e. it is independent of the value of the variable Z. If in a linear-in-parameters model without interaction   the variable X and the variable Z influence Y but X does not influence Z, the total effect of X on Y is equal to the direct effect of X on Y. This is not true in a loglinear model without interaction:  I find, indeed, that when $\mu^{X=1,Z=1,Y=1}=1$, the total effect is not equal to the direct effect, but there is another effect, which I call cell effect. The cell effect is present only if more variables influence the same variable, as in this case where X and Z influence Y. The cell effect formula is:

\begin{equation}
\begin{split}
\mbox{Cell}^{\mbox{effect}}_{x,x'} (Z)=& \left [\frac{\sum_z P(Y|X=x',Z=z) P(Z|X=x)}{1-\sum_z P(Y|X=x',Z=z) P(Z|X=x)} \right.\\
& \left. \frac{1- \sum_z  P(Y|X=x, Z=z) P(Z|X=x)}{\sum_z P(Y|X=x, Z=z) P(Z|X=x)} \right] \\
& \left [\frac{P(Y|X=x',Z=z)}{1-P(Y|X=x',Z=z)} \frac{1-P(Y|X=x,Z=z)}{ P(Y|X=x,Z=z)} \right]^{-1}
\end{split}
\end{equation}
\\
It is not linked to the interaction calculated in additive form. The additive interaction in a loglinear model  it is obtain by this formula:

\begin{equation*}
\pi^{Y=1|X=1,Z=1}-\pi^{Y=1|X=0,Z=1}-\pi^{Y=1|X=1,Z=0}+\pi^{Y=1|X=0,Z=0}
\end{equation*}

In a loglinear model without the multiplicative interaction   with dummy code, the additive interaction effect  is linked to linearity (appendix A) and is equal to 0 in these 3 cases:  in the first case if the two-effects parameter between Y ad X is equal to 1 (i.e. $\mu^{X=1,Y=1}=1$), in the second case if the two-effects parameter between Y and Z is equal to 1 (i.e. $\mu^{Z=1,Y=1}=1$) and in the third case if the two-effects parameter between Y and Z is equal to $ (\mu^{Y=1})^{-2} (\mu^{X=1,Y=1})^{-1}$. Of course when there is the multiplicative interaction, the additive interaction exists.

In a loglinear model with dummy code and without multiplicative interaction, the cell effect is equal to

\begin{equation}
\begin{split}
\mbox{Cell}^{\mbox{effect}}_{x=0,x'=1}= &\frac{\eta^{Y|X=0,Z=0}+ \eta^{Y|X=0, Z=1} \mu_c^{Z=1}}{\eta^{Y|X=0,Z=0}+ \eta^{Y|X=0, Z=1} \mu_c^{Z=1} \mu^{Z=1,Y=1}}\\
& \frac{\eta^{Y|X=1,Z=0}+ \eta^{Y|X=1, Z=1} \mu_c^{Z=1} \mu^{Z=1,Y=1}}{\eta^{Y|X=1,Z=0}+ \eta^{Y|X=1, Z=1} \mu_c^{Z=1} } \\
\end{split}
\end{equation}
\\
Of course, if the parameter $\mu^{Z=1,Y=1}$ is equal to 1 or $\mu^{X=1,Y=1}$ is equal to 1, the cell effect becomes equal to 1 and the total effect is equal to the direct effect of X on Y or of Z on Y. In this case the cell effect depends on $Z|X$ and then I can write  $\mbox{Cell}^{\mbox{effect}}_{x,x'} (Z)= \mbox{Cell}^{\mbox{effect}}_{x,x'}$, i.e. the cell effect can be interpreted as a constant interaction effect (this is not true in a loglinear model with multiplicative interaction). As seen in the introduction, indeed, the interaction effect can cause that the direct effect of one variable on another is a function of a third variable, and therefore varies as the third variable varies,  while in this case the cell effect remains constant as the third variable varies.

Because the total effect and the direct effect used in the loglinear literature are the odds ratio versions of the total effect and the controlled direct effect proposed by \cite{Pearl01,Pearl09,Pearl12}, then I propose the odds ratio version of his indirect effect:

\begin{equation}
\begin{split}
OR_{x,x'} ^{IE} = & \frac{\sum_z P(Y|X=x,Z=z) P(Z|X=x')}{1-\sum_z P(Y|X=x,Z=z) P(Z|X=x')}\\ &\frac{1-\sum_z P(Y|X=x,Z=z) P(Z|X=x)}{\sum_z P(Y|X=x,Z=z) P(Z|X=x)}
\end{split}
\end{equation}
\\
Then the total effect is equal to

\begin{equation}
OR^{TE}_{x,x'}=OR_{x,x'}^{LDE}(z)\mbox{Cell}_{x,x'}^{\mbox{effect}} \frac{1}{OR_{x',x} ^{IE}} 
\end{equation}
\\
The direct effect used in the loglinear literature and the cell effect form the odds ratio version of Pearl's natural direct effect. Pearl, indeed, proposes 2 direct effects: the natural direct effect and the controlled direct effect. The first  is the change of Y when X changes and Z is constant at whatever value obtained by the start value of X, while the second is the change of Y  when X  changes and all  other factors are held fixed. The natural direct effect is:

\begin{equation}
\begin{split}
OR_{x,x'}^{NDE} & = OR_{x,x'}^{LDE}(Z)\mbox{Cell}_{x,x'}^{\mbox{effect}}(Z)\\
& =\frac{\sum_z P(Y|X=x',Z=z) P(Z|X=x)}{1-\sum_z P(Y|X=x',Z=z) P(Z|X=x)}\frac{1- P(Y|X=x)}{P(Y|X=x)}\\
\end{split}
\end{equation}
\\
The natural direct effect depends on $Z|X$  for Pearl's definition 

 The interpretation of the effects calculated as odds ratio is the following: a value of the effect bigger than 1 means that the 2 variables change in the same direction  (if X increases, Y increases) and a value of the effect smaller than 1 means that the 2 variables change in the different direction (if X increases, Y decreases).  

If I want calculate the effects of the variation of X from x' a x, I obtain

\begin{equation*}
OR^{TE}_{x',x}= \frac{1}{ OR^{TE}_{x,x'}}
\end{equation*}

\begin{equation*}
OR^{LDE}_{x',x}=\frac{1}{ OR^{LDE}_{x,x'}}
\end{equation*}

\begin{equation*}
OR^{NDE}_{x',x} \neq \frac{1}{ OR^{NDE}_{x,x'}}
\end{equation*}

\begin{equation*}
OR^{IE}_{x',x} \neq \frac{1}{ OR^{IE}_{x,x'}}
\end{equation*}

\begin{equation*}
\mbox{Cell}^{\mbox{effect}}_{x',x} \neq  \frac{1}{\mbox{Cell}^{\mbox{effect}}_{x,x'}}
\end{equation*}

Now I consider the relation among the effects and the parameters. In literature, the causal two-effects parameters ($\mu^{X=1, Y=1},\mu^{Z=1,Y=1},\mu_c^{X=1, Z=1}$) determine the presence or absence of the direct link  between the variables: for example if I suppose that $\mu^{X=1, Y=1}$ (recalling that in this case $\mu_c^{X=1,Y=1}=\mu^{X=1,Y=1}$) is equal to 1, then  there is not a direct effect of X on Y. In terms of path diagram, the arrow which goes from X to Y is not present. If I set the  causal two-effects parameter $\mu_c^{X=1, Z=1}$ equal to 1, I eliminate the direct effect of X on Z, while if I  set the no causal two-effects parameter $\mu^{X=1, Z=1}$ equal to 1, I  don't eliminate the direct effect of X on Z, this because only the causal parameters can determine the presence or absence of the direct link. This can be shown using a simple example. I consider the following no causal parameters: $\mu^{X=1,Y=1}=0.02$, $\mu^{Z=1,Y=1}=0.01$, $\mu^{Y=1}$=0.2, $\mu^{X=1,Z=1}=1$, $\mu^{Z=1}=2$ and $\mu^{X=1}=1.5$. The no causal parameter $\mu^{ X=1,Z=1}$ is equal to 1, i.e. there is not a  effect between $Z$ and $X$. If I calculate the indirect effect, I find that $OR^{IE}$ is equal to 0.8894, i.e. an effect mediated by Z exists. This occurs because  $\mu_c^{ X=1,Z=1}$ is equal to 1.1929 and then the variable X is still linked causally to Z, also if the no causal parameter is equal to 1.  The total effect $OR^{TE}$ is equal to $OR^{LDE}$ because the cell effect is equal to the inverse of the indirect effect $OR^{IE}$ which measures the inverse change of X (from $x'$ to $x$).

 Now I consider a new loglinear model where the values of parameters  $\mu^{XY}, \mu^{ZY}, \mu^{Y}, \mu^{Z}$ and $ \mu^{X}$ remain equal to those of the previous example and  the value of $\mu^{X=1,Z=1}$  becomes 0.8383. In this case $OR^{IE}$ is equal to 1 because the causal parameter  $\mu_c^{X=1,Z=1}$ is equal to 1. 
 In conclusion, if $\mu^{X=1,Z=1}$ or $\mu_c^{X=1,Z=1}$ is equal to 1, the total effect $OR^{TE}$ is equal to the direct effect used in the loglinear literature $OR^{LDE}$ or to the natural direct effect, but in the first case there is the indirect effect, while in the second case, it disappears. When there is not the indirect effect, the variable X influences Y only directly.

Now I consider a causal loglinear model with the multiplicative interaction. Then Y is influenced directly by the variable X, by variable Z and by their joint effect due to the three-interaction term. Using the definition of multiplicative interaction, the direct effect of X on Y used in the loglinear literature becomes a function of Z.  I show this recalling that the formulas (4), (5), (8) and (10) remain valid and  applying the formula (5) to a causal loglinear model with dummy code. The direct effect used in the loglinear literature becomes:

\begin{equation*}
OR^{LDE}_{x=0,x'=1} (Z) = \mu^{X=1,Y=1} \mu^{X=1, Z=z, Y=1}
\end{equation*}
\\
For the same reason,  also the cell effect becomes a function of Z:

\begin{equation*}
\begin{split}
\mbox{Cell}^{\mbox{effect}}_{x=0,x'=1} (Z)= &\frac{1}{ \mu^{X=1,Z=z,Y=1}}\frac{\eta^{Y|X=0,Z=0}+ \eta^{Y|X=0, Z=1} \mu_c^{Z=1}}{\eta^{Y|X=0,Z=0}+ \eta^{Y|X=0, Z=1} \mu_c^{Z=1} \mu^{Z=1,Y=1}} \\
&\frac{\eta^{Y|X=1,Z=0}+ \eta^{Y|X=1, Z=1} \mu^{X=1,Z=1,Y=1} \mu_c^{Z=1} \mu^{Z=1,Y=1}}{\eta^{Y|X=1,Z=0}+ \eta^{Y|X=1, Z=1} \mu_c^{Z=1} } \\
\end{split}
\end{equation*}
\\
The natural direct, indirect and total effects, instead, do not become function of Z. The indirect effect of a model with multiplicative interaction remains equal to that of a model without multiplicative interaction.

\section{efflog package}
\subsection{Estimation procedure}
In the first section, I have presented two formulations of the same model: they are founded on two different assumptions (causal model and no causal model) and are estimated with two different approaches.  In a loglinear model without the multiplicative interaction the  parameters of the additive form can be estimated so:

\begin{verbatim}
#Loglinear model:
fit.glm<-glm(count~.^2, data=table, family=poisson)
# where table is the frequency of the variables X,Z and Y
\end{verbatim}
while in a causal loglinear model without the multiplicative interaction, I use the package efflog \cite{gloria} to estimate the parameters of the additive form

\begin{verbatim}
# Causal loglinear model:
\library(efflog)
Cloglin(table)
# where table is the frequency of the variables X,Z and Y
\end{verbatim}
Of course to obtain the parameters of the multiplicative form, it is sufficiently to make this transformation $\mu= exp(log(\mu))$. In efflog there is the command  \begin{verbatim}exp_par(table) \end{verbatim} which calculates the causal parameters in multiplicative form. The parameters of the causal form (i.e those with subscript c) are estimated by the causal loglinear model, the parameters without subscript are estimated by the traditional loglinear model. Only the parameters of conditional probability $\pi^{Y=y|X=x, Z=z}$ remain equal in both forms and for this reason I do not use never the subscript c for them.

Now I consider that the three-interaction $\mu^{X=1,Z=1,Y=1}$ is different from 1. Then the path diagram of the only direct effects on Y is shown in Figure 2. The introduction of the three-interaction parameter produces a multiplicative interaction effect on Y. If I consider the marginal probability of X and the conditional probabilities of Z and of Y, the introduction of the interaction term modifies only the formula  (3) : now  the three-interaction term is added to the conditional probability of Y given X and Z so  the model becomes:

\begin{equation*}
\pi^{X=x}= \eta_c^{X} \mu_c^{X=x}
\end{equation*}
\begin{equation*}
\pi^{Z=z| X=x}= \eta_c^{Z|X=x} \mu_c^{Z=z}  \mu_c^{  X=x,Z=z}
\end{equation*}
\begin{equation*}
\pi^{Y=y|X=x, Z=z}=\eta^{Y| X=x, Z=z} \mu^{Y=y} \mu^{X=x,Y=y } \mu^{ Z=z,Y=y} \mu^{X=x,Z=z,Y=y}
\end{equation*}
Then in a loglinear model with the multiplicative interaction the  parameters of the additive form can be estimated so:

\begin{verbatim}
#Loglinear model:
fit.glm<-glm(count~.^3, data=table, family=poisson)
# where table is the frequency of the variables X,Z and Y
\end{verbatim}
while in a causal loglinear model with the multiplicative interaction, I use the package efflog \cite{gloria}

\begin{verbatim}
# Causal loglinear model:
\library(efflog)
Cloglin_mult(table)
# where table is the frequency of the variables X,Z and Y
\end{verbatim}
Of course to obtain the parameters of the multiplicative form, it is sufficiently to make this transformation $\mu= exp(log(\mu))$. In efflog there is the command \begin{verbatim}exp_par_mult(table) \end{verbatim} which calculates the causal parameters in multiplicative form.

\subsection{Causal effects}
The package efflog \cite{gloria} provides functions to calculate directly these effects. The commands for the effects of a loglinear model without multiplicative interaction are:

\begin{verbatim}
cell_effect_or(x,y,z,w) 
ndirect_effect_or(x,y,z,w,t)
indirect_effect_or(x,y,z,w,t)
total_effect_or(x,y,z,w,t)
\end{verbatim}
where $x= \mu^{Y=y}$, $y=\mu^{X=x,Y=y}, z=\mu^{Z=z,Y=y}, w=\mu_c^{Z=z}, t=\mu_c^{X=x,Z=z}$.

The commands for calculating the effects of a loglinear
model with multiplicative interaction are:

\begin{verbatim}
cell_effect_mult_or(x,y,z,w,q)
ndirect_effect_mult_or(x,y,z,w,t,q)
indirect_effect_or(x,y,z,w,t)
total_effect_mult_or(x,y,z,w,t,q)
\end{verbatim}
where $x= \mu^{Y=y}$, $y=\mu^{X=x, Y=y,}, z=\mu^{Z=z, Y=y}, w=\mu_c^{Z=z}, t=\mu_c^{X=x, Z=z}, q=\mu^{X=x,Z=z, Y=y}$.

\section{Empirical examples}
In this section, I apply my causal theory and the package efflog to empirical results. They consider the relations between a typical product (in this case the Sauris' ham) and its festival. This analysis is developed in marketing but it can be applied in many economic fields or in social sciences.

\subsection{Example 1}

The first dataset is composed of 3  dichotomous variables (X measures the interest about Sauris' ham considering the possibility of buying Sauris'  ham, Z measures the satisfaction about Sauris' festival considering the happiness which  an individual  has if he thinks about Sauris' festival and Y measures the future behavior considering if an individual will buy Sauris's ham more often). The results of the causal loglinear model are shown in table 5.  The two-effects parameters are all significant (i.e. all are different from 1).   According to the traditional loglinear literature, the causal two-effects parameters are the direct effect. In this case, because all causal two-effects parameters are greater than 1, then an increase  of variable X produces an increase of variable Z and the same result occurs for the relation between X and Y and for that between Z and Y.   Now I calculate the effects using the formulas (4), (6), (8) and (10).  The total effect is equal to 2.4008, then an increase of X produces an increase of Y, the natural direct effect is equal to 1.8741, then  an increase of X produces an increase of Y.  The indirect effect  is equal to 1.2845: an increase of X produces , indirectly, an increase of Y. The cell effect is 0.9741: it mitigates the controlled direct effect. The presence of 2 variables which influence Y causes the cell effect and then the natural direct effect becomes 1.8741. Now I control if the additive interaction effect is equal to 0, i.e. if $\mu^{Z=1,Y=1}=(\mu^{Y=1})^{-2} (\mu^{X=1,Y=1})^{-1}$. 
 I apply a z-test (appendix B)  and I find that  z is equal to 0.4174  (the p-value is 0.6764) then I accept the hypothesis that the additive interaction effect is equal to 0. This is an example where the additive interaction is equal to 0 and cell effect is different from 1.

From this analysis, I conclude that if a customer becomes interested in Sauris' ham, then he will buy  Sauris' ham more often also thanks to the happiness due to Sauris' festival. In marketing research, this means that a event linked to the product can increase its sell. However, the role of this event is minus important than the interest about the product (indirect effect/ total effect $<$ direct effect/total effect ) and their joint effect decreases  the  direct effect used in the loglinear literature (cell effect $<$ 1).

\begin{table}[t]
\begin{minipage}[c]{.4\textwidth}
\centering
\begin{tabular}{||l|l||}
\hline
parameter & value  \\
$\mu^{XY}$ & 	1.9240** \\
$\mu^{ZY}$ & 	2.4038***\\
$\mu^{Y}$ &	  0.4881***\\
$\mu_c^{XZ}$ &	  3.3059***\\
$\mu_c^{Z}$ &	 0.4659***\\
$\mu_c^{X}$ &	1.7132***\\
& \\
\hline
\end{tabular}
\caption{First dataset}
\end{minipage}
\begin{minipage}[c]{.4\textwidth}
\centering
\begin{tabular}{||l|l||}
\hline
parameter & value  \\
$\mu^{XZY}$ & 2.8826* \\
$\mu^{XY}$ & 	1.4042 \\
$\mu^{ZY}$ & 	3.5385**\\
$\mu^{Y}$ &	   0.2826***\\
$\mu_c^{XZ}$ &	 3.5534*** \\
$\mu_c^{Z}$ &	0.3390***  \\
$\mu_c^{X}$ &	1.2278.\\
\hline
\end{tabular}
\caption{Second dataset}
\end{minipage}
\caption*{\footnotesize{Signif. codes:  0 "***" 0.001 "**" 0.01 "*" 0.05 "." 0.1 " " 1 }}
\end{table}

\subsection{Example 2}

Now I consider a second dataset. This dataset is composed of 3 dichotomous variables (X measures the interest about Sauris' ham considering the possibility of testing Sauris'  ham, Z measures the satisfaction about Sauris' festival considering the  quality of products presented during the Sauris' festival and Y measures the future behavior considering if an individual will suggest others to go to Sauris' festival).  The  values of parameters are shown in table 6. The total effect is equal to 3.1886, then an increase of X produces an increase of Y. The natural direct effect is equal to 1.7286, then  an increase of X produces an increase of Y.  The indirect effect  is equal to 1.4493: an increase of X produces, indirectly, an increase of Y. The cell effect, i.e. the effect of  the presence of 2 variables which influence future behavior,  is 0.4270 with Z=1, i.e. it mitigates the LD effect, while the cell effect with Z=0 is 1.231002, i.e. it  increases the LD effect. Now I consider the effect of the multiplicative interaction, whose parameter is bigger than 1. When Z is high (Z=1),  the joint effect of satisfaction and interest (multiplicative interaction effect) increases the positive LD effect, while when it is low (Z=0), it leaves intact the LD effect.  For any value of satisfaction, then, the overall interaction effect (cell effect $+$ multiplicative interaction effect) is positive because $\mu^{X=1,Y=1}$ is equal to 1.4 and the natural direct effect is always equal to 1.7286. 

From this analysis, I conclude that if a customer becomes interested in Sauris' ham, then he will suggest to go  Sauris' festival more often thanks also to the quality of the presented products and to the overall joint effect of interest and of satisfaction.

\section{Summary}

When a researcher analyzes the data, he is interested in understanding the mechanisms which govern
the changes of the variables. To understand these mechanisms he uses the causal effects. Unfortunately,
when the researcher uses the loglinear models to study the data, he has not available a causal theory,
but only few comments on various papers where the odds ratios are used. For this reason, using
the causal concepts provided by \cite{Pearl01,Pearl09,Pearl12}, I provide a r-package efflog (\cite{gloria}) to calculate the effects in the loglinear models using odds ratios so that the parameters have the same interpretation given
by the loglinear literature. Making so I find a new effect which I call 
cell effect. It can be interpreted as an interaction effect which occurs whenever I consider two
variables affecting a third. The interaction effects in a causal loglinear model are three: multiplicative interaction effect, additive interaction effect and cell effect. Then the researcher, who studies his data with the causal theory proposed in this paper and using the r-package efflog will have the traditional effects (direct, indirect and total) plus a new interaction effect. 

\bibliography{gheno}

\appendix
\section{Appendix A: Additive interaction: a measure of the linearity}

 I analyze the relation among the causal log-linear model and the linearity. As seen in the section 2, the causal log-linear model analyzes the relation among the variables using the cell frequencies, then the causal relation can be analyzed only using the conditional probabilities, but the relation among the variables can be expressed by any function, for example $Y=f(X)$. A simple linear model requires that the causal relation among the variables is linear, i.e $Y= \beta_0+ \gamma_1 X$. Of course the world is not perfect: it is necessary to introduce an error term, then the relation between Y and X becomes $Y= \beta_0 + \gamma_1 X + \zeta$. In its simpler formulation, the linear model considers  the variables $X, Y$ and $\zeta$ continuous and  normally distributed. Now I analyze what occurs in a log-linear model if the relation between X and Y is linear. In a first step, I consider a perfect word, i.e. where Y is perfectly given by the relation $\beta_0 + \gamma_1 X$ and X and Y are continuous variables with a generic joint distribution $P(X,Y)$. To analyze the same variables with a causal log-linear model, I must discretize the continuous variables. Now I transform X and Y in two binary variables $X^*$ and $Y^*$ so: the the values of X (or Y) which are smaller than the mean become 0, the values of X (or Y) which are bigger than the mean become 1. This particular transformation is made in order that the linearity is inserted in causal log-linear model. The marginal probabilities of the new variables $X^*$ and $Y^*$ are:

\begin{equation}
P(X^*)=\left\{ \begin{array} {cc}
  
P(X< E(X)) &  X^*=0  \\
P(X \geq E(X)) & X^*=1 \\
\end{array} \right.
\end{equation}

\begin{equation}
P(Y^*)=\left\{ \begin{array} {cc}
  
P(Y< E(Y)) &  Y^*=0  \\
P(Y \geq E(Y)) & Y^*=1 \\
\end{array} \right.
\end{equation}
\\
\\
Now, using the linear relation between X and Y, I obtain that :

\begin{equation*}
Y < E(Y) => \beta_0 + \gamma_1 X < \beta_0 + \gamma_1 E(X)
\end{equation*}
\\
\\
I simplify and obtain that

\begin{equation*}
Y < E(Y)  \quad \mbox{is equal to} \quad \gamma_1 X < \gamma_1 E(X)
\end{equation*}
\\
\\
i.e

\begin{equation}
P(Y^*=0)=P(Y<E(Y))=\left\{ \begin{array} {cc}
  
P(X< E(X))=P(X^*=0) &  \mbox{if} \quad \gamma_1>0  \\
P(X> E(X)) &  \mbox{if} \quad \gamma_1 <0 \\
\end{array} \right.
\end{equation}
\\
\\
To analyze the relation between $X^*$ and $Y^*$, I consider the variables T and W, which are so built:

\begin{equation*}
W=\left\{ \begin{array} {cc}
 1 & \mbox{if} \quad P(Y^*=1|X^*=1) \\
 0 & \mbox{if} \quad P(Y^*=0|X^*=1) \\
\end{array} \right.
\end{equation*}

\begin{equation*}
T=\left\{ \begin{array} {cc}
 1 & \mbox{if} \quad P(Y^*=0|X^*=0) \\
 0 & \mbox{if} \quad P(Y^*=1|X^*=0) \\
\end{array} \right.
\end{equation*}

If $\gamma_1$ is positive, the joint distribution of T and W is showed in table 7: without error, the probability of $Y^*$ equal to 1  given $X^*$ equal to 1 is 1, i.e. it is the certain event. The event "$Y^*$ equal to 0 given $X^*$ equal to 0" is the certain event.

\begin{table}[t]
\begin{minipage}[b]{0.4\textwidth}
\centering
\begin{tabular}{||l|l|ll|l||}
\hline
& & W && \\
&& 1 & 0 & \\
\hline
T & 1 & 1 & 0 & 1 \\
   & 0 & 0 & 0 & 0 \\
\hline
   &    & 1 & 0   & 1 \\
\hline
\end{tabular}
\caption{The joint probability with $\gamma_1 >0$, without error term}
\end{minipage}
\hspace{20 pt}
\begin{minipage}[b]{0.4\textwidth}
\centering
\begin{tabular}{||l|l|ll|l||}
\hline
& & W && \\
&& 1 & 0 & \\
\hline
T & 1 & $\pi_{11}$ & $\pi_{10}$ & $\pi_{1+}$ \\
   & 0 & $\pi_{01}$ & $\pi_{00}$ & $\pi_{0+}$ \\
\hline
   &    & $\pi_{+1}$ &  $\pi_{+0}$   & 1 \\
\hline
\end{tabular}
\caption{The joint probability with $\gamma_1 >0$ and error term}
\end{minipage}
\end{table}

Because X is  a continuous variable, the sign of equality in the inequality is not important, then the formula (13) can be written so:

\begin{equation}
P(Y^*=0)=P(Y<E(Y))=\left\{ \begin{array} {cc}
  
P(X< E(X))=P(X^*=0) &  \mbox{if} \quad \gamma_1>0  \\
P(X> E(X))= P(X^*=1) &  \mbox{if} \quad \gamma_1 <0 \\
\end{array} \right.
\end{equation}
\\	
Unfortunately, the world is not perfect and the relation between X and Y contains an error term, which has zero mean. Then I obtain:

\begin{equation*}
Y < E(Y) => \beta_0 + \gamma_1 X + \zeta < \beta_0 + \gamma_1 E(X)
\end{equation*}
\\
I simplify and obtain that

\begin{equation*}
Y < E(Y)= \gamma_1 E(X) \quad \mbox{is equal to} \quad \gamma_1 [X- E(X)] < - \zeta
\end{equation*}

With error term and $\gamma_1$ bigger than 0, the joint probability of variables T and W is showed in table 8: there is not the certain event as the case without error term because the presence of the error term produces the existence of  discordant events (i.e. "$Y^*$ equal to 0 given $X^*$ equal to 1" or "$Y^*$ equal to 1 given $X^*$ equal to 0"). I follow the Tutz's method (\cite{Tu}) for  the repeated measurements for binary variables. The repeated measurements occur when the researcher measures the same variables at different time or under different conditions. To analyze if the distribution changes over times or conditions, he considers the joint distribution of the repeated measurements and controls if the marginal homogeneity holds. The marginal homogeneity can be seen in the table 7: it holds if $\pi_{+1}$ is equal to $\pi_{1+}$. In the perfect world, the marginal homogeneity calculated for the joint distribution of the binary variables T and W holds: $\pi_{+1}=1=\pi_{1+}=1$. In the imperfect world the homogeneity holds iff $\pi_{+1}$ is equal to $\pi_{1+}$. I consider the log linear model showed in table 9, where I use the dummy code. Then the marginal homogeneity condition becomes:

\begin{equation*}
P(Y^*=1|X^*=1)= \frac{ \mu^{Y^*=1} \mu^{X^*=1} \mu^{Y^*=1,X^*=1} } {\mu^{X^*=1} [ 1+ \mu^{Y^*=1} \mu^{Y^*=1,X^*=1}]} = P(Y^*=0|X^*=0)=  \frac{ 1}{ 1+ \mu^{Y^*=1}}
\end{equation*}

i.e. the two "not" causal parameter $\mu^{Y^*=1,X^*=1}$ is equal to reciprocal of the  squared "not" causal parameter $\mu^{Y=1}$ (i.e. $\mu^{Y^*=1,X^*=1}= 1/ [\mu^{Y^*=1}]^2)$.

Now I consider a mediation linear model in a perfect world, where X influences linearly Y and Z, which influences in turn linearly Y. This model is so:

\begin{equation*}
Z= \alpha_0 + \alpha_{1} X
\end{equation*}

\begin{equation*}
Y= \omega_0 + \omega_{1} X + \omega_2 Z
\end{equation*}
\\
This model can be rewritten in reduced form, i.e:

\begin{equation*}
Y= (\omega_0 + \alpha_0) + (\omega_1 + \omega_2 \alpha_1 )X = \beta_0 + \gamma_1 X
\end{equation*}
\\
which is equal to  the relation between X and Y analyzed until now.  Now I transform X, Z and Y in binary variables $X^*$, $Z^*$ and $Y^*$   ( 0 if the value of variable is smaller than its mean, 1 if the value of variable is bigger than its mean). As in the simple linear model if $\alpha_1$ is positive and there is not error term ,  the probability $P(Z^*=0)$ is equal to probability  $P(X^*=0)$. Now If $\omega_1$ , $\alpha_1$ and $\omega_2$ are positive, also the probability $P(Y^*=0)$ is equal to probability $P(X^*=0)$, because in this case in reduce form $\gamma_1$ is positive. Then the variables W and T becomes:
\\
\\
\begin{equation*}
W=\left\{ \begin{array} {cc}
 1 & \mbox{if} \quad P(Y^*=1|X^*=1,Z^*=1) \\
 0 & \mbox{if} \quad P(Y^*=0|X^*=1;Z^*=1) \\
\end{array} \right.
\end{equation*}

\begin{equation*}
T=\left\{ \begin{array} {cc}
 1 & \mbox{if} \quad P(Y^*=0|X^*=0, Z^*=0) \\
 0 & \mbox{if} \quad P(Y^*=1|X^*=0, Z^*=0) \\
\end{array} \right.
\end{equation*}
\\
\\
\begin{table}[t]
\begin{minipage}[b]{0.4\textwidth}
\centering
\begin{tabular}{||l|l|l||}
\hline
\footnotesize{$x^*$}  & \footnotesize{$y^*$} & $\pi^{X^*=x^*,Y^*=y^*}$ \\
\hline
\footnotesize{0} & \footnotesize{0} & \footnotesize{$\eta$}  \\
\footnotesize{0} & \footnotesize {1}& \footnotesize {$\eta \mu^{Y^*=1}$ }\\
\footnotesize{1} & \footnotesize{0} & \footnotesize{$\eta \mu^{X^*=1}$ }\\
\footnotesize{1} & \footnotesize{1} & \footnotesize{$ \eta \mu^{Y^*=1} \mu^{X^*=1} \mu^{Y^*=1,X^*=1}$ } \\
\hline
\end{tabular}
\caption{The joint probability of simple linear model}
\end{minipage}
\hspace{ 2 pt}
\begin{minipage}[b]{0.5\textwidth}
\centering
\begin{tabular}{||l|l|l|l||}
\hline
\footnotesize{$x^*$} & \footnotesize{$z^*$} & \footnotesize{$y^*$} & $\pi^{X^*=x^*,Z^*=z^*,Y^*=y^*}$ \\
\hline
\footnotesize{0} & \footnotesize{0} & \footnotesize{0} & \footnotesize{$\eta$  }\\
\footnotesize{0} & \footnotesize{0} & \footnotesize{1} & \footnotesize{$\eta \mu^{Y^*=1}$ } \\
\footnotesize{0} & \footnotesize{1} & \footnotesize{0} & \footnotesize{$\eta \mu^{Z^*=1}$}  \\
\footnotesize{0} & \footnotesize{1} & \footnotesize{1} & \footnotesize{$\eta \mu^{Y^*=1} \mu^{Z^*=1} \mu^{Y^*=1,Z^*=1}$ } \\
\footnotesize{1} & \footnotesize{0} & \footnotesize{0} & \footnotesize{$\eta \mu^{X^*=1}$}  \\
\footnotesize{1} & \footnotesize{0} & \footnotesize{1} & \footnotesize{$\eta \mu^{X^*=1} \mu^{Y^*=1} \mu^{Y^*=1,X^*=1}$}  \\
\footnotesize{1} & \footnotesize{1} & \footnotesize{0} & \footnotesize{$\eta \mu^{X^*=1} \mu^{Z^*=1} \mu^{X^*=1,Z^*=1}$ } \\
\footnotesize{1} & \footnotesize{1} & \footnotesize{1} &  \footnotesize{$ \eta \mu^{X^*=1} \mu^{Y^*=1} \mu^{Z^*=1} \mu^{X^*=1,Y^*=1}$ }\\
   &    &    &  \footnotesize{$\mu^{X^*=1,Z^*=1} \mu^{Y^*=1, Z=1} \mu^{X^*=1,Y^*=1, Z^*=1}$}\\
\hline
\end{tabular}
\caption{The joint probability of mediation linear model }
\end{minipage}
\end{table}

In this case, the conditional probabilities  $Y^*$ given $Z^*$ and $X^*$ are all equal to 0 in a perfect world when  $Y^*=X^*=Z^*=1$ and $Y^*=X^*=Z^*=0$ do not occur. The log-linear model is showed in table 10. If  I introduce the error terms and I use Tutz's method, of course with multiplicative interaction parameter $\mu^{Y=1,X=1,Z=1}$  equal to 1, the marginal homogeneity condition becomes:

\begin{equation*}
\frac{ \mu^{Y^*=1} \mu^{Y^*=1,Z^*=1} \mu^{Y^*=1,X^*=1}}{ 1+\mu^{Y^*=1} \mu^{Y^*=1,Z^*=1} \mu^{Y^*=1,X^*=1}}= \frac{  1}{ 1+\mu^{Y^*=1} }
\end{equation*}
\\
i.e. the "two" not causal parameter $\mu^{Y^*=1,X^*=1}$ is equal to product between the two not causal parameter $\mu^{Y^*=1,Z^*=1}$ and reciprocal of the  squared  "not" causal parameter $\mu^{Y=1}$ (i.e $\mu^{Y^*=1,X^*=1}= $ $1/  \{[\mu^{Y^*=1}]^2$ $ \mu^{Y^*=1, Z^*=1} \}$. This condition doesn't imply the condition $P(Y^*=1|X^*=1) =P(Y^*=0|X^*=0)$: indeed the error term  in the relation between the variable $X^*$ and $Z^*$ causes the inequality  $P(Y^*=1|X^*=1) \neq P(Y^*=0|X^*=0)$. Then I must consider also the relation between the variable $X^*$ and $Z^*$. The marginal homogeneity condition holds iff $\mu_c^{Z^*=1,X^*=1}= 1/ (\mu_c^{Z^*=1})^2$, where c defines that the parameters are those of a causal log-linear model. As seen in section 2, the causal parameters can be always transformed in not causal log-linear parameters. Then the mediation linear model implies that:

\begin{equation}
\begin{array}{c}
\mu^{Y^*=1,X^*=1}= \frac{1} {[\mu^{Y^*=1}]^2 \mu^{Y^*=1, Z^*=1} }\\
\mu_c^{Z^*=1,X^*=1}= \frac{1}{ (\mu_c^{Z^*=1})^2}
\end{array} 
\end{equation}
 
If these two conditions are satisfied, the equivalence $P(Y^*=1|X^*=1) =P(Y^*=0|X^*=0)$ is true. Then I conclude that if I suppose that the variables X,Z and Y are linearly linked, then the relative parameters of the causal model must satisfy the bonds (15). This is important because the first bond of (15) causes the nullity of the additive interaction effect in a causal loglinear model without multiplicative interaction effect.

\section{Appendix B: Test on the presence of the additive interaction term in a loglinear model without interaction}

In this section I find a test to analyze the presence of the cell effect obtained by Pearl's causal formula. For simplicity, I consider the parameters of the additive form. \cite{Agresti} shows that the no causal loglinear model without two-effects parameter for a 2x2 table (i.e. a contingency table for 2 variable, X and Y)  can be so written:

\begin{equation}
\log (\boldsymbol{m})=
\left[
\begin{array}{c}
\log m^{X=0,Y=0}\\
\log m^{X=0,Y=1}\\
\log m^{X=1,Y=0}\\
\log m^{X=1,Y=1}\\
\end{array}
\right]
=
\left[
\begin{array}{ccc}
1&0&0 \\
1& 0& 1\\
1& 1& 0 \\
1& 1 & 1 \\
\end{array}
\right]
\left[
\begin{array}{c}
\log(\eta) \\
\log (\mu^{X})\\
\log (\mu^{Y}) \\
\end{array}
\right ]
= \boldsymbol{D} \boldsymbol{\lambda}
\end{equation}
where $\boldsymbol{m}$ denotes the column vector of the expected counts of the contingency table and  $\boldsymbol{ \lambda}$ is the vector of the additive no causal log-linear parameters. The formula (16) can be extent to a loglinear model with all interactions for a nxn contigency table.  Then, in a general no causal log-linear model, using the maximum likelihood method, the variance-covariance matrix for the estimated additive no causal parameters is

\begin{equation}
Cov(\log ( \boldsymbol{\hat{\mu}})) =Cov( \boldsymbol{\hat{\lambda}})=[ \boldsymbol{D}' diag (\boldsymbol{\hat{m}})\boldsymbol{ D}]^{-1}
\end{equation}

Now I consider the particular case where the additive interaction is equal to 0 also if 2 variables influence the variable Y . This occurs  if $\mu^{Y=1,Z=1}=[(\mu^{Y=1})^2 \mu^{X=1,Y=1}]^{-1}$. Because these parameters remain equal  both in the causal loglinear model and in the loglinear model, this relation can be tested both in the causal loglinear model and in the loglinear model. I test this relation in the loglinear model. For simplicity, I consider the additive parametrization, then the relation becomes:

\begin{equation}
\begin{split}
\log(\mu^{Z=1,Y=1}) + 2 \log(\mu^{Y=1})+& \log( \mu^{X=1,Y=1})=\\
& \lambda^{Z=1,Y=1} + 2\lambda^{Y=1} + \lambda^{X=1,Y=1}=0
\end{split}
\end{equation}

Now I propose a z-test.  Because the vector of the estimated lambda are distributed as a multivariate normal, the left-side (18) is a variable normally distributed with the mean equal to 

\begin{equation*}
\begin{split}
E(\hat{\lambda}^{Z=1,Y=1} & +  2\hat{\lambda}^{Y=1} + \hat{\lambda}^{X=1,Y=1}) =\\
& E( \hat{\beta}),\\
\end{split}
\end{equation*}
i.e. 
\begin{equation*} 
\lambda^{Z=1,Y=1} + 2\lambda^{Y=1} + \lambda^{X=1,Y=1}= \beta,
\end{equation*}
and the variance equal to 
\begin{equation*}
\begin{split}
Var(\hat{\beta})= & Var( \hat{\lambda}^{Z=1,Y=1}) \\
& + 4 Var(\hat{\lambda}^{Y=1}) + Var(\hat{\lambda}^{X=1,Y=1})+ 4 Cov(\hat{\lambda}^{Z=1,Y=1},\hat{\lambda}^{Y=1})  \\
&
 + 2 Cov( \hat{\lambda}^{Z=1,Y=1},\hat{\lambda}^{X=1,Y=1})\\
& + 4 Cov (\hat{\lambda}^{X=1,Y=1}, \hat{\lambda}^{Y=1})\\
\end{split}
\end{equation*}
\\ 
Then  the statistic z, which is equal to $(\hat{\beta} - \beta) ( Var(\hat{\beta}))^{-1/2}$, is normally distributed with mean equal to 0 and variance equal to 1. Now I has a statistic to test when the cell effect is equal to 0. The equality (18) requires that $\beta$ is equal to 0, then testing the equality condition is equal to testing that  $\beta$ is equal to 0.

\end{article}

\end{document}